# Sessile liquid drop evaporation: analytical solution in bipolar coordinates

Peter Lebedev-Stepanov

FSRC "Crystallography and Photonics" RAS, Leninskii pr-t, 59, Moscow, 119333, Russia

National Research Nuclear University MEPhI, Kashirskoye shosse, 31, Moscow, 115409, Russia

Electronic mail: petrls@yandex.ru

April, 6, 2021

## Abstract

The investigation of evaporating liquid drop deposited onto a flat surface is of great importance for physical, engineering and medical applications. Novel analytical expressions are proposed to calculate the evaporation rate of sessile drop (mass loss per unit surface area per unit time) and total evaporation rate (mass loss per unit time). To obtain these results, the H.M. Macdonald's solution for a flat wedge was transformed by method of inversion in a sphere originally developed by J.C. Maxwell in the Treatise on Electricity and Magnetism with further derivation the solution for a lens based on consideration given in bipolar coordinates by G.A. Grinberg. These solutions are mathematically equivalent to expressions proposed earlier by Yuri O. Popov [Phys. Rev. E **71**, 036313 (2005)], but, in some cases, probably, the new solutions can be more useful from a computational point of view.

## I. INTRODUCTION

Evaporating liquid drop deposited onto a flat surface is an important object both for theoretical modeling (evaporation dynamics, microfluidics inside the drop, dissolved nanoparticle dynamics in the volume of the evaporating drop, nanoparticle ensemble self-assembly, etc.), Refs. [1-4] and applications (printing technologies, functionalized coatings, medical diagnostics etc.), Refs. [5-10]. There are three basic problems to estimate the colloidal particles self-assembly in evaporating droplets, Ref. [11]: 1) solute evaporation from the droplet surface to surrounding air (outer problem), 2) hydrodynamic flows in droplet volume (inner problem), 3) particle dynamics into droplet with account of interparticle interactions, particle-surfaces interactions, particle-flow interactions, solvation effects. The first of these problems, evaporation, has a primary importance due to it is an origin driving force of self-assembly process.

It is easy to derive that the equilibrium shape of a sessile drop of a slowly evaporating liquid, the size of which is much smaller than the capillary constant (Bond number, Bo<<1), approximately corresponds to a spherical segment with the given contact angle. If evaporation is controlled by the diffusion of vapor from the drop surface into the surrounding air at a constant temperature, then the quasi-stationary solution can be obtained from the Laplace equation

$$\nabla^2 n = 0, \tag{1}$$

where *n* is a vapor volume concentration outside the drop with the boundary conditions $n = n_S$ at the drop air-liquid surface and $n = n_\infty$ far from the drop.

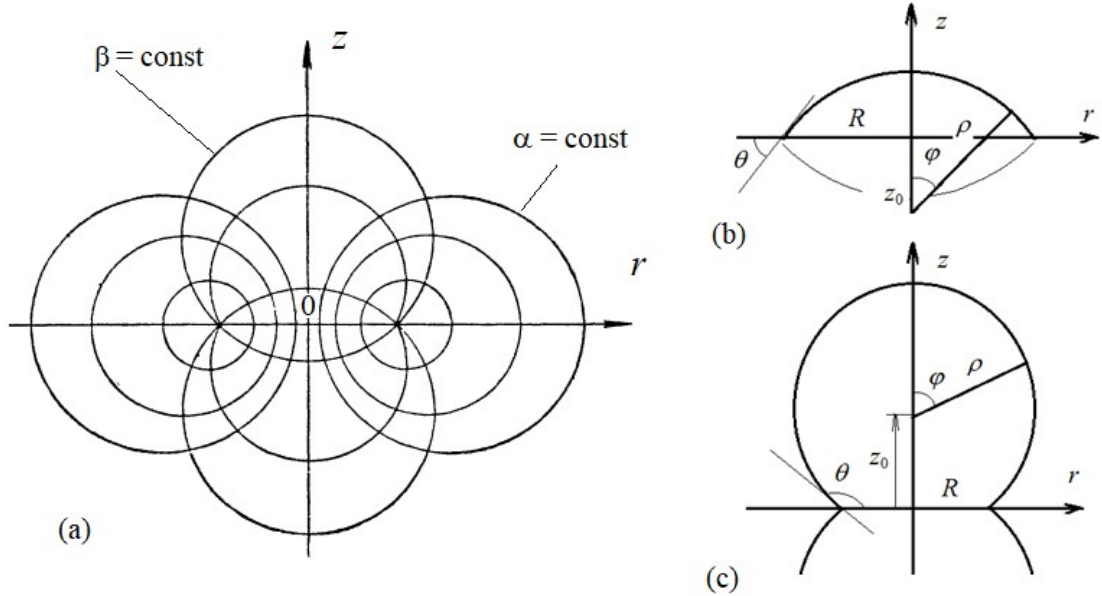

**FIG.1.** (a) Toroidal coordinates $(\alpha, \beta)$. Geometry of drops with acute (b) and obtuse (c) contact angles.

Using a general solution proposed in the monograph by N.N. Lebedev [12], Deegan et al. [1], Hu and Larson [13], Popov [2] obtained the expression for the vapor concentration in space near a sessile drop in toroidal coordinates $(\alpha, \beta)$ [Fig.1(a)]

$$n(\alpha,\beta) = n_\infty + (n_s - n_\infty)\sqrt{2(\cosh\alpha - \cos\beta)} \int_0^\infty \frac{\cosh\theta\tau \cosh(2\pi-\beta)\tau}{\cosh\pi\tau \cosh(\pi-\theta)\tau} P_{-1/2+i\tau}(\cosh\alpha) d\tau, \quad (2)$$

where $\theta$ is the contact angle of the drop [Fig.1(b),(c)], $P_{-1/2+i\tau}(\cosh\alpha) = \frac{2}{\pi}\cosh\pi\tau \int_0^\infty \frac{\cos\tau t}{\sqrt{2(\cosh t + \cosh\alpha)}} dt$ are the Legendre functions of the first kind.

The toroidal coordinates $(\alpha, \beta)$ are related to the cylindrical coordinates *r* and *z* by [Fig.1(b) and (c)]

$$r = \frac{R\sinh\alpha}{\cosh\alpha - \cos\beta}, \quad z = \frac{R\sin\beta}{\cosh\alpha - \cos\beta}. \quad (3)$$

Note that in the case of a drop with an obtuse angle [Fig.1(c)], representation given by Eq. (3) is not unambiguous. There can be two $z$ values for the same radius $r$. Therefore, it is more convenient to use polar coordinates $(\rho, \varphi)$ rather than cylindrical ones $(r, z)$.

The evaporation rate (evaporative flow density) from the surface of the drop is presented in Ref. [2]

$$J(\alpha) = D \frac{n_s - n_\infty}{R} \times \left[ \frac{\sin\theta}{2} + \sqrt{2}(\cosh\alpha + \cos\theta)^{3/2} \int_0^\infty \frac{\cosh\theta\tau}{\cosh\pi\tau} \tanh[(\pi-\theta)\tau] P_{-1/2+i\tau}(\cosh\alpha)\tau d\tau \right], \quad (4)$$

where $D$ is a diffusion coefficient of the vapor in the air. Here, toroidal coordinate $\alpha$ ranges in the interval from 0 (top of the drop) to $\infty$ (contact line).

There are two basic conventional solutions for the total evaporation rate of sessile drop (mass loss per unit time). The first was given by Picknett and Bexon in Ref. [14], and the other was developed by Popov in Ref. [2]. Popov's expression is given by

$$W = \pi R D(n_s - n_\infty) \left[ \frac{\sin\theta}{1+\cos\theta} + 4 \int_0^\infty \frac{1+\cosh 2\theta\tau}{\sinh 2\pi\tau} \tanh[(\pi-\theta)\tau] d\tau \right]. \quad (5)$$

Picknett and Bexon relied on an exact solution obtained from the electrostatic analogy of problem (1). They derived the following expression in Ref. [8]:

$$W = 2\pi D(n_s - n_\infty)C = 2\pi\rho D(n_s - n_\infty)g(\theta), \quad (6)$$

where «capacity» of drop, $C = g(\theta)\rho$, is calculated by approximation formulas

$$\begin{aligned} g(\theta) &= 0.6366\theta + 0.09591\theta^2 - 0.06144\theta^3, \quad 0 \le \theta \le 0.175; \\ g(\theta) &= 0.00008957 + 0.6333\theta + 0.1160\theta^2 - 0.08878\theta^3 + 0.01033\theta^4, \quad 0.175 \le \theta \le \pi. \end{aligned} \quad (7)$$

The approximation error of the exact analytical expression does not exceed 0.2%: It can be shown that Eqs. (5) and (6) give an equivalent result and make it possible to calculate the evaporation time of a droplet for any dependence the droplet radius on contact angle, $R(\theta)$.

Using the limits, it can be derived that the evaporation rate of a disk of radius $R$ is described by the formula $W = 4DR(n_s - n_\infty)$; in the case of hemispherical drop ($\theta = 0.5\pi$), it gives $W = 2\pi D(n_s - n_\infty)\rho$; in the case of the sphere deposited to flat surface ($\theta = \pi$), we obtain $W = 4\pi D(n_s - n_\infty)\rho \ln 2$ ($\rho$ is the spherical segment radius corresponded to drop surface [Fig.1(b),(c)]).

In the following part of this work, the derivation of alternative expressions of evaporation rate of sessile drop (mass loss per unit surface area per unit time), total evaporation rate (mass loss per unit time) and vapor concentration obtained in bipolar coordinates is presented.

## II. DERIVATION OF DROP EVAPORATION RATE

Let us consider another solution of the problem (1) using an electrostatic analogy by the inversion method originally developed by J.C. Maxwell, Ref. [15], that allows to convert the solution of conducting wedge problem to the solution that corresponds to conducting lens. This way was considered in detail by G.A. Grinberg, Ref. [16].

Let AO and BO be two intersecting conducting planes that are perpendicular to the plane of Fig. 2. They form a wedge with an internal angle $2\theta$.

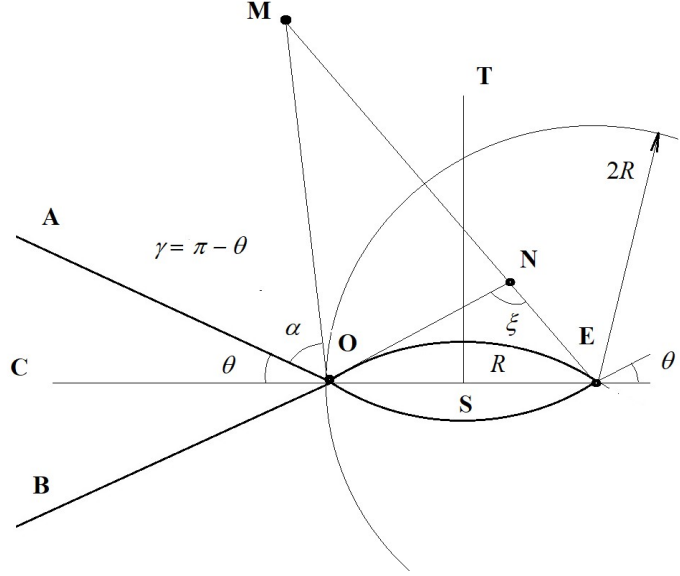

**FIG.2.** Conducting wedge AOB that transforms into conducting lens by inversion.

A charge $e$ is placed at a point E located at a distance of $2R$ from the edge of the wedge O on the extension of the bisector of the angle AOB. There is a point M characterized by cylindrical coordinates $(r, \alpha, 0)$ which is centered at point O, where $r = |\text{OM}|$. The polar angle $\alpha$ is measured from the ray OA to the right (Fig.2). According to Macdonald's formula, Ref. [17], this charge creates a potential at the point $(r, \alpha, z)$ determined by

$$\varphi(r,\alpha,z) = \frac{e}{4\gamma\sqrt{Rr}} \int_{\eta}^{\infty} \left[ \frac{\sinh\frac{\pi\varsigma}{2\gamma}}{\cosh\frac{\pi\varsigma}{2\gamma} - \cos\frac{\pi(\alpha-\gamma)}{2\gamma}} - \frac{\sinh\frac{\pi\varsigma}{2\gamma}}{\cosh\frac{\pi\varsigma}{2\gamma} - \cos\frac{\pi(\alpha+\gamma)}{2\gamma}} \right] \frac{d\varsigma}{\sqrt{\cosh\varsigma - \cosh\eta}}, \quad (8)$$

where $\cosh\eta = \dfrac{4R^2 + r^2 + z^2}{4Rr}$. Taking into account $\alpha = \gamma - \xi$ and $\cos\dfrac{\pi(\alpha-\gamma)}{2\gamma} = \pm\sin\dfrac{\pi\alpha}{2\gamma}$, the expression (8) can be rewritten as

$$\varphi(r,\alpha,z) = \frac{e}{2\gamma\sqrt{Rr}} \sin\frac{\pi\alpha}{2\gamma} \int_{\eta}^{\infty} \sinh\frac{\pi\varsigma}{2\gamma} \left( \cosh^2\frac{\pi\varsigma}{2\gamma} - \sin^2\frac{\pi\alpha}{2\gamma} \right)^{-1} \frac{d\varsigma}{\sqrt{\cosh\varsigma - \cosh\eta}}. \quad (9)$$

It was assumed that the wedge has zero potential at the surface.

The inversion about the center at point E for a sphere of radius 2R transforms the wedge into a lens, the geometry of which is shown in Fig. 2. The potential of the reflected system (equipotential lens) at the reflected point N with coordinates ($r'$, $\alpha'$, 0), where $r' = |ON|$, and $\alpha'$ is the angle AON, is determined by

$$\psi(r',\alpha',0) = -\frac{e}{2R} + \frac{2R}{|EN|}\varphi(r,\alpha,0) . \qquad (10)$$

By definition,

$$|EN| \cdot |EM| = 4R^2 , \qquad (11)$$

where 2R is the radius of the sphere in which the point M is reflected to N.

Taking into account Eq. (9), we get

$$\psi(r',\alpha',0) = V - \frac{V}{|EN|}\frac{2R^{1.5}}{\gamma\sqrt{r}}\sin\frac{\pi\alpha}{2\gamma}\int_{\eta}^{\infty}\sinh\frac{\pi\varsigma}{2\gamma}\left(\cosh^2\frac{\pi\varsigma}{2\gamma} - \sin^2\frac{\pi\alpha}{2\gamma}\right)^{-1}\frac{d\varsigma}{\sqrt{\cosh\varsigma - \cosh\eta}} , \qquad (12)$$

where $V = -\frac{e}{2R}$.

We introduce bipolar coordinates ($\xi$, $\omega$) centered at points O and E. By definition,

$$\omega = \ln\frac{|ON|}{|EN|}. \qquad (13)$$

Eq. (11) can be rewritten as

$$\frac{|EN|}{2R} = \frac{2R}{|EM|}. \qquad (14)$$

It follows that the triangles OME and NOE are similar: one angle is common, and two sides are proportional. Then we obtain

$$\angle MOE = \xi, \quad \alpha = \gamma - \xi \text{ and } r = 2Re^{\omega} . \qquad (15)$$

The cosine theorem gives

$$\frac{1}{|EN|^2 r} = \frac{\cosh\omega - \cos\xi}{4R^3}. \qquad (16)$$

If $z = 0$, we obtain $\cosh\eta = \frac{4R^2 + r^2}{4Rr} = \cosh\rho$, therefore $\eta = \pm\omega$. $\qquad (17)$

Taking into account the above equations, we derive

$$\psi(\omega,\xi) = V - \frac{2V}{\gamma}(\cosh\omega - \cos\xi)^{1/2}\cos\frac{\pi\xi}{2\gamma}\int_{\omega}^{\infty}\left(\cosh\frac{\pi\varsigma}{\gamma} - \cos\frac{\pi\xi}{\gamma}\right)^{-1}\frac{\sinh\frac{\pi\varsigma}{2\gamma}d\varsigma}{\sqrt{\cosh\varsigma - \cosh\omega}} . \qquad (18)$$

The charge density on the lens surface is given by

$$\sigma = -\frac{\cosh\omega - \cos\xi}{R}\frac{\partial\psi(\omega,\xi)}{\partial\xi}. \qquad (19)$$

In a bipolar coordinate system, the spherical segment shown in Fig. 2 is characterized by the angle $\xi$. This angle has the same value for all points on the circular arc OE with a given contact angle $\theta$ (Fig.3)). Considering the isosceles triangle OQS (Fig. 3), it is easy to derive that

$$\xi = \gamma, \text{ where } \gamma = \pi - \theta . \qquad (20)$$

Taking into account the Eqs. (18)-(20), we get

$$\sigma(\omega) = \frac{\pi V}{2(\pi-\theta)^2 R}(\cosh\omega - \cos(\pi-\theta))^{3/2} \int_\omega^\infty \frac{\sinh\frac{\pi\varsigma}{2(\pi-\theta)}}{\cosh^2\frac{\pi\varsigma}{2(\pi-\theta)}} \frac{d\varsigma}{\sqrt{\cosh\varsigma - \cosh\omega}} \qquad (21)$$

or

$$\sigma(\omega) = \frac{V}{(\theta-\pi)R}(\cosh\omega - \cos(\pi-\theta))^{3/2} \int_{\varsigma=\omega}^\infty \frac{d\,\text{sech}\frac{\pi\varsigma}{2(\pi-\theta)}}{\sqrt{\cosh\varsigma - \cosh\omega}}, \qquad (22)$$

where
$$\omega = \ln\frac{|OP|}{|EP|}. \qquad (23)$$

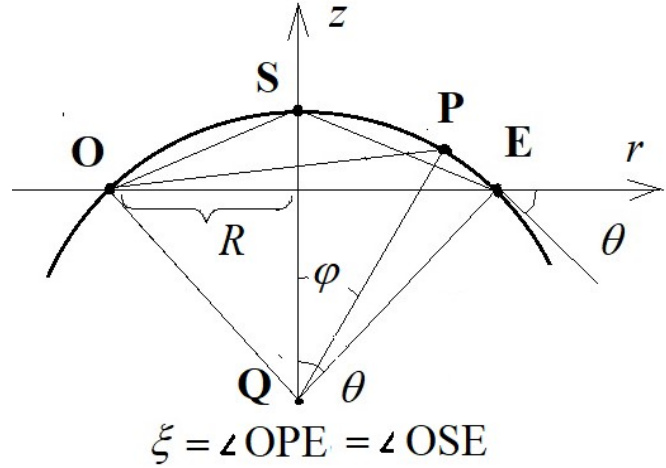

$\xi = \angle OPE = \angle OSE$

**FIG.3.** Geometry of the spherical segment corresponded to the top part of the lens with given contact angle $\theta$ that is shown in Fig.2

The following geometric relationships with parameters $\varphi$, $\omega$ and $\xi$ takes place (see Fig.3):

$$r = \frac{R\sinh\omega}{\cosh\omega - \cos\xi}, \qquad (24)$$

$$z = \frac{R\sin\xi}{\cosh\omega - \cos\xi}, \qquad (25)$$

$$\cosh\omega = \frac{\sin^2\theta}{\cos\varphi - \cos\theta} - \cos\theta, \qquad (26)$$

$$\omega(\varphi) = \frac{1}{2}\ln\frac{1-\cos(\theta+\varphi)}{1-\cos(\theta-\varphi)}, \text{ where } \varphi \in [0,\theta]. \qquad (27)$$

Using Eqs. (24) - (27), one can find $\omega$ corresponding to any angular position $\varphi$ of a point P on a segment surface. Then, the two-dimensional charge density at a given point on the segment surface can be determined by Eq. (21) (Fig.3). By this way, we can derive the explicit dependence $\sigma$ on $\varphi$ (Fig.1 (b),(c)):

$$\sigma(\varphi) = \frac{\pi V}{(\pi-\theta)^2 R} \frac{\sin^3\theta}{\cos\varphi - \cos\theta} \int_\varphi^\theta \frac{(1-\cos(\theta+\beta))^{\frac{\pi}{2(\pi-\theta)}} - (1-\cos(\theta-\beta))^{\frac{\pi}{2(\pi-\theta)}}}{\left((1-\cos(\theta+\beta))^{\frac{\pi}{2(\pi-\theta)}} + (1-\cos(\theta-\beta))^{\frac{\pi}{2(\pi-\theta)}}\right)^2} \frac{(\cos\beta - \cos\theta)^{\frac{\pi}{2(\pi-\theta)} - \frac{1}{2}} d\beta}{\sqrt{\cos\varphi - \cos\beta}}$$

(28)

The total surface charge is

$$Q = 2\pi R^2 \int_1^\infty \frac{\sigma \sinh\omega}{(\cosh\omega - \cos(\pi-\theta))^2} d\omega,$$

(29)

hence

$$Q = \frac{\pi^2 RV}{(\pi-\theta)^2} \int_{\omega=1}^\infty \frac{\sinh\omega}{\sqrt{\cosh\omega - \cos(\pi-\theta)}} \left[\int_\omega^\infty \frac{\sinh\frac{\pi\varsigma}{2(\pi-\theta)}}{\cosh^2\frac{\pi\varsigma}{2(\pi-\theta)}} \frac{d\varsigma}{\sqrt{\cosh\varsigma - \cosh\omega}}\right] d\omega.$$

(30)

Using the analogy of the problem of sessile drop evaporation with the electrostatics problem presented here, we can adopt the Eq. (18) to the vapor density

$$n(\omega,\xi) = n_S - \frac{2(n_S - n_\infty)}{\pi-\theta}(\cosh\omega - \cos\xi)^{1/2} \cos\frac{\pi\xi}{2(\pi-\theta)} \int_{\varsigma=\omega}^\infty \left(\cosh\frac{\pi\varsigma}{(\pi-\theta)} - \cos\frac{\pi\xi}{(\pi-\theta)}\right)^{-1} \frac{\sinh\frac{\pi\varsigma}{2(\pi-\theta)} d\varsigma}{\sqrt{\cosh\varsigma - \cosh\omega}}.$$

(31)

Analogously, the evaporation rate (mass loss per unit surface area per unit time) can be written taking into account Eqs. (21), (22) and (28)

$$J(\omega) = \frac{D(n_S - n_\infty)}{(\pi-\theta)R}(\cosh\omega - \cos(\pi-\theta))^{3/2} \int_{\varsigma=\omega}^\infty \frac{d\operatorname{sech}\frac{\pi\varsigma}{2(\pi-\theta)}}{\sqrt{\cosh\varsigma - \cosh\omega}}$$

(32)

or

$$J(\varphi) = \frac{\pi D(n_S - n_\infty)}{(\pi-\theta)^2 R} \frac{\sin^3\theta}{\cos\varphi - \cos\theta} \int_\varphi^\theta \frac{(1-\cos(\theta+\beta))^{\frac{\pi}{2(\pi-\theta)}} - (1-\cos(\theta-\beta))^{\frac{\pi}{2(\pi-\theta)}}}{\left((1-\cos(\theta+\beta))^{\frac{\pi}{2(\pi-\theta)}} + (1-\cos(\theta-\beta))^{\frac{\pi}{2(\pi-\theta)}}\right)^2} \frac{(\cos\beta - \cos\theta)^{\frac{\pi}{2(\pi-\theta)} - \frac{1}{2}} d\beta}{\sqrt{\cos\varphi - \cos\beta}}.$$

(33)

The mass loss per unit time is

$$W = \frac{\pi^2 R D(n_S - n_\infty)}{(\pi - \theta)^2} \int_0^\infty \frac{\sinh\omega}{\sqrt{\cosh\omega - \cos(\pi - \theta)}} \left[ \int_\omega^\infty \frac{\sinh\frac{\pi\varsigma}{2(\pi-\theta)}}{\cosh^2\frac{\pi\varsigma}{2(\pi-\theta)}} \frac{d\varsigma}{\sqrt{\cosh\varsigma - \cosh\omega}} \right] d\omega \ . \quad (34)$$

Eq. (33) can be rewritten as

$$J(\varphi) = \frac{D(n_S - n_\infty)}{R} f(\varphi), \quad (35)$$

where

$$f(\varphi) = \frac{\pi}{(\pi-\theta)^2} \frac{\sin^3\theta}{\cos\varphi - \cos\theta} \int_\varphi^\theta \frac{(1-\cos(\theta+\beta))^{\frac{\pi}{2(\pi-\theta)}} - (1-\cos(\theta-\beta))^{\frac{\pi}{2(\pi-\theta)}}}{\left( (1-\cos(\theta+\beta))^{\frac{\pi}{2(\pi-\theta)}} + (1-\cos(\theta-\beta))^{\frac{\pi}{2(\pi-\theta)}} \right)^2} \frac{(\cos\beta - \cos\theta)^{\frac{\pi}{2(\pi-\theta)} - \frac{1}{2}} d\beta}{\sqrt{\cos\varphi - \cos\beta}}. \quad (36)$$

The graphs of the function given by Eq.(36) for the different values of contact angles are shown in the.Fig.4. The behavior of this function presents the dependence of evaporation flow density on polar angle $\varphi$ that is denoted in Fig. 1(b) and (c). In appearance, this is the same dependence $J(\varphi)$ that was reported in the Refs. [1, 2, 13], but this is given by a different formal representation.

In case of $\theta \to 0$ (lubrication approximation), Eq.(36) gives

$$f(\varphi) \approx \frac{2\theta}{\pi} (\theta^2 - \varphi^2)^{-\frac{1}{2}}. \quad (37)$$

or

$$J(r) = \frac{2D(n_s - n_\infty)}{\pi R} \left( 1 - \frac{r^2}{R^2} \right)^{-1/2} \quad (38)$$

with $r \approx \rho\varphi$ and $R \approx \rho\theta$ where $\rho$ and $R$ are shown in Fig.1, $r$ is the cylindrical coordinate on the drop surface which corresponds to given $\varphi$.

Integrating the flow Eq.(38) over a circle of radius $R$, one can obtain the total evaporation rate

$$W = 4DR(n_s - n_\infty). \quad (39)$$

Result (39) was also obtained in the different ways by Picknett and Baxon in Ref. [14], by Fuchs in Ref. [18]. Result given by Eq. (38) was obtained earlier by Deegan et al (Ref.[1]) and Hu and Larson from the Eq.(4).

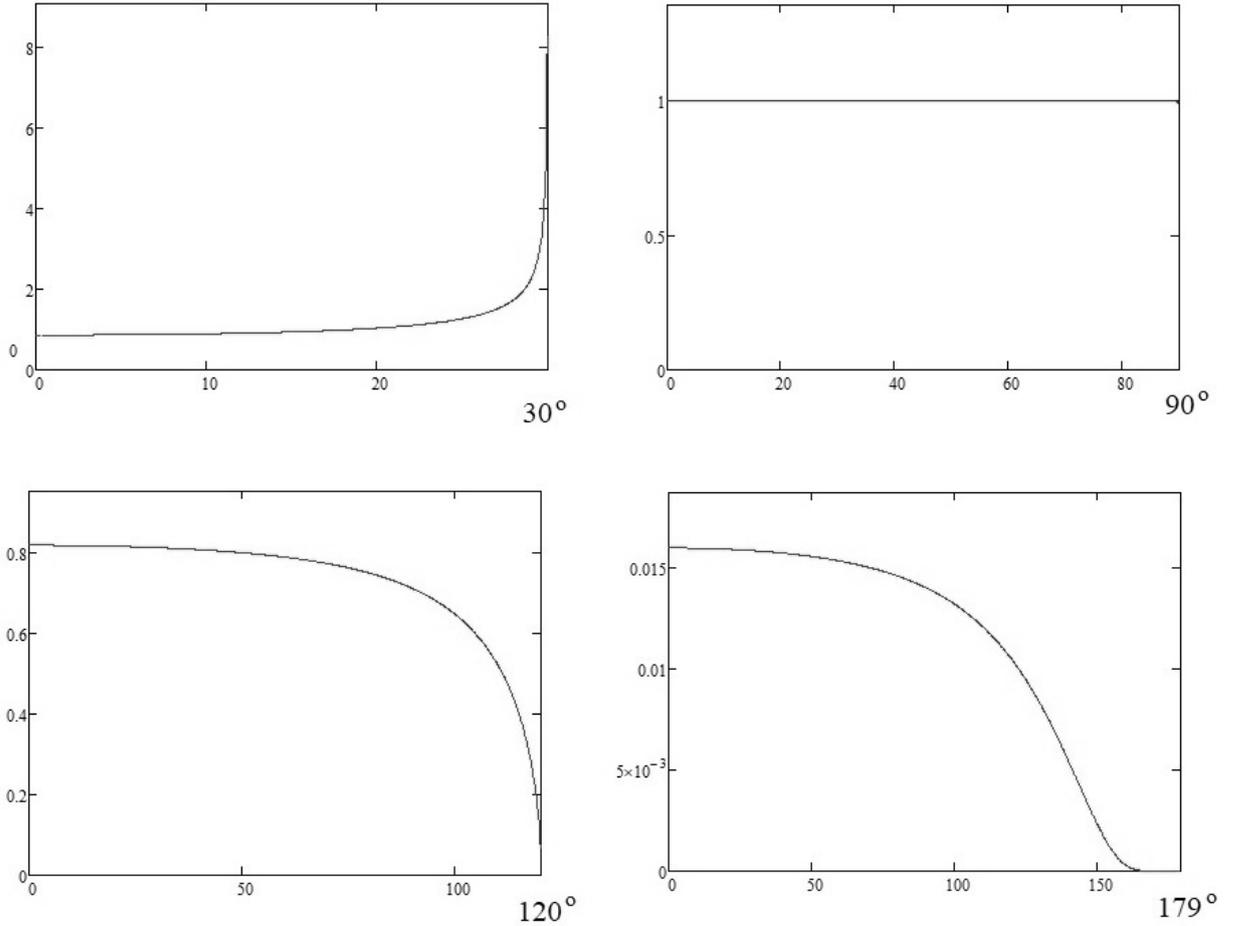

**FIG.4.** The graphs of the function $f(\varphi)$ given by Eq. (36) for the different values of contact angles of the drop ($\theta = 30°, 90°, 120°, 179°$).

## III. CONCLUSION

Novel analytical expressions of evaporation rate of sessile drop (mass loss per unit surface area per unit time) and total evaporation rate (mass loss per unit time) are found. To obtain these results, the H.M. Macdonald's solution (Ref. [17]) for a flat wedge was transformed by method of inversion in a sphere originally developed by J.C. Maxwell (Ref. [15]) with further derivation the solution for a lens based on consideration given by G.A. Grinberg in bipolar coordinates (Ref. [16]).

These solutions are mathematically equivalent to expressions proposed earlier by Yuri O. Popov [2], but, in some cases, probably, our solutions have more simple form and can be more useful from a computational point of view. So, Eqs. (32) and (33) for the flow density $J$ are represented in the form of the single integrals, while corresponding Popov's expression given by Eq. (4) is actually a double integral taking into account that the function $P_{-1/2+i\tau}(\cosh\alpha)$ has an integral representation. Therefore, we guess that the new expressions for the flow density proposed in this work are more convenient in calculations of $J$ than the "old" formula Eq.(4). On the other hand, the Popov's expression (5) for the integral mass loss $W$ has the form of a single integral that can be more useful in calculation than the corresponding formula given by Eq.(34) obtained in this work in the form of a double integral.

## ACKNOWLEDGMENTS

The work was supported by the "Improving of the competitiveness" program of the National Research Nuclear University MEPhIThe and by the Russian Foundation for Basic Researches grant no. 19-02-00937.

______________________________________________